\documentstyle[11pt,twocolumn,epsfig]{article}

\begin{document}
\begin{small}
\title{\large \bf Quantum correlations and classical resonances in an open chaotic
system}


\author{\small Wentao T. Lu,
Kristi Pance,
Prabhakar Pradhan
and S. Sridhar\footnote{e-mail: srinivas@neu.edu}\\
\small Physics Department, Northeastern University, Boston,
Massachusetts 02115.}


\maketitle

\begin{footnotesize}

\begin{abstract}
We show that the autocorrelation of quantum spectra 
of an open chaotic system is well described by the classical Ruelle-Pollicott
 resonances of the associated chaotic strange repeller. 
This correspondence is demonstrated utilizing microwave experiments on 
2-D $n$-disk billiard geometries, by determination of the wave-vector 
autocorrelation $C(\kappa )$ from the experimental quantum spectra $S_{21} (k)$. 
The correspondence is also established via ``numerical experiments" that 
simulate $S_{21} (k)$ and $C(\kappa )$ using periodic orbit calculations
 of the quantum and classical resonances. Semiclassical arguments that relate 
quantum and classical correlation functions in terms of fluctuations of 
the density of states and correlations of particle density are also examined
and support the experimental results. The results establish a 
correspondence between quantum spectral correlations and classical 
decay modes in an open systems.
\end{abstract}

\end{footnotesize}


Pacs ref: 05.45.Mt, 05.45.Ac, 03.65.Sq, 84.40.-x


\section{Introduction}

The correspondence between quantum and classical mechanics has turned out to
be particularly rich for systems that are classically chaotic \cite
{Gutzwiller90}. This correspondence is usually considered in terms of the
asymptotic behavior of quantum properties. Statistical analysis of the
quantum spectra of closed chaotic systems display short-range correlations
that are universal and can be described by random matrix theory (RMT) \cite
{Bohigas90,Guhr98}. The long range behavior of spectral fluctuations require
semiclassical treatments that incorporate periodic orbits\cite{Berry91}.

In this paper we establish a new connection between the statistics of
quantum spectra and classical dynamical properties of an open chaotic
system, utilizing physical and numerical experiments on the well-known $n$%
-disk billiard geometry. The $n$-disk system \cite{Gaspard89} is a paradigm
of open hyperbolic systems, with an associated strange repeller \cite
{Kadanoff} whose strange properties arise from the Cantor set of the
unstable periodic orbits (PO). For hyperbolic repellers, Ruelle \cite
{Ruelle86} and Pollicott \cite{Pollicott} showed that the dynamical
evolution of repellers toward equilibrium can be expressed in terms of
complex poles, leading to the so-called Ruelle-Pollicott (RP) resonances.
The $n$-disk billiards have been extensively addressed theoretically, both
classically and semiclassically \cite
{Gaspard89,Gaspard92,Cvitanovic93,Gaspard94}, and only recently have been
studied experimentally \cite{Lu99,Lu00}.

Our microwave experiments on the $n$-disk billiards measure the quantum
transmission spectrum $S_{21}(k)$, which is essentially the two-point
Green's function $G({\vec{r}}_{1},{\vec{r}}_{2},k)$, and from which the
spectral autocorrelation $C(\kappa )$ is determined. The small wave vector $%
\kappa $ (long time) behavior of the spectral autocorrelation provides a
measure of the quantum escape rate, and is in good agreement with the
corresponding classical escape rate $\gamma _{cl}$ \cite{Lu99,Lu00}, as
predicted by RMT analysis of this universal behavior. For large $\kappa
>\gamma _{cl}$ (short time), non-universal oscillations of the
autocorrelation $C(\kappa )$ are observed, that can be understood completely
in terms of classical RP resonances. This behavior is similar to that of the
variance $\Sigma ^{2}(L)\,$of closed chaotic systems, which also displays
leading universal behavior for small $L$, followed by non-universal behavior
for large $L$ that can be described in terms of system specific periodic
orbits \cite{Berry91}. The present work demonstrates that $C(\kappa )$ of
open systems displays a similar trend from short-range universality to
long-range non-universality, however the non-universal contributions to
quantum correlations are described in terms of other classical phase space
structures of the associated repeller, viz. the RP resonances \cite{ponote}.

A concise account of the main results was presented in Ref. \cite{Pance00}.
This paper further explores the details of the experimental aspects, as well
as confirming the results via numerical experiments using simulated quantum
spectra. Semiclassical arguments that demonstrate this correspondence are
discussed in terms of fluctuations of the density of states and correlations
of the particle density. The wider implications of the present results are
also discussed.

\section{Experimental results}

The quantum dynamics of the 2-D $n$-disk system can be realized in a
microwave experiment exploiting the mapping between the Helmholtz equation
and the Schr\"{o}dinger equation in their stationary forms and in the 2-D
limit. This is because under the conditions of the experiment, the
Maxwell-Helmholtz equations reduce to $(\nabla ^{2}+k^{2})\Psi =0$ with $%
\Psi =E_{z}$ the microwave electric field. The experimental geometry
consists of thin copper disks sandwiched between two large copper plates of
size $55\times 55$ $cm^{2}$ in area. In order to simulate an infinite
system, microwave absorber material ECCOSORB AN-77 was placed between the
plates at the edges. Microwaves were coupled in and out through antennas
inserted in the vicinity of the scatterers. All measurements were carried
out using an HP8510B vector network analyzer which measured the complex
transmission amplitude ($S_{21}$) and reflection amplitude ($S_{11}$) $S$%
-parameters of the coax + scatterer system. An example of the measured
transmission $T(k)=|S_{21}(k)|^{2}$ of a 4-disk system is shown in Fig. 1.
For more details of the microwave experiments, see Ref. \cite{Lu00}.

An open system can be represented by an effective Hamiltonian consisting of
two parts, $H=H_{c}+iW$, where $H_{c}$ is the Hamiltonian for the closed
system and $W=W^{\dagger }$ represents the decay to open channels. Since the
total Hamiltonian is not Hermitian, the eigenfunctions $\psi _{n}$ of $H$
with eigenvalues $E_{n}$ do not form an orthogonal set. Instead, let $\phi
_{n}$ be the eigenfunction of the adjoint operator $H^{\dagger }$ with
eigenvalues $E_{n}^{*}$, and the $\psi $'s and $\phi $'s form a
bi-orthonormal set with $\int \psi _{m}^{*}({\vec{r}})\phi _{n}({\vec{r}})d{%
\vec{r}=\delta }_{mn}.$ The Green's function for an open system is \cite
{Datta} $G({\vec{r}}_{1},{\vec{r}}_{2},k)=\sum_{n}\frac{\psi _{n}^{*}({\vec{r%
}}_{1})\phi _{n}({\vec{r}}_{2})}{E-E_{n}}$. For the $n$-disk system, with
the quantum resonances in wave vector space $k_{n}=s_{n}+is_{n}^{\prime }$,
we have

\begin{eqnarray*}
|G({\vec{r}}_{1},{\vec{r}}_{2},k)|^{2}\hskip -.3cm &=&\hskip -.3cm
\frac{2m}{\hbar ^{2}}\sum_{n}\frac{%
|\psi _{n}({\vec{r}}_{1})\phi _{n}({\vec{r}}_{2})|^{2}}{%
[(k+s_{n})^{2}+s_{n}^{\prime 2}][(k-s_{n})^{2}+s_{n}^{\prime 2}]} \\
\hskip -.3cm&&\hskip -.3cm+\frac{2m}{\hbar ^{2}}\sum_{n,m}\,^{^{\prime }}
\frac{\psi _{n}^{*}({\vec{r}%
}_{1})\phi _{n}({\vec{r}}_{2})\psi _{m}({\vec{r}}_{1})\phi _{m}^{*}({\vec{r}}%
_{2})}{(k^{2}-k_{n}^{2})(k^{2}-k_{m}^{2})^{*}}.
\end{eqnarray*}
The second sum with the prime includes the off-diagonal terms with $n\neq m$%
. Their contribution can be neglected if all the resonances are well
separated.

The transmission amplitude $S_{21}(k)$ measured between two point antennas
is essentially the Green's function $G({\vec{r}}_{1},{\vec{r}}_{2},k)$ of
the system, so that $S_{21}(k)=A(k)G({\vec{r}}_{1},{\vec{r}}_{2},k)$ with $%
A(k)$ a slowly-varying modulating function \cite{Lu00}. Setting $%
c_{n}=(2m/\hbar ^{2})|A(k)\psi _{n}({\vec{r}}_{1})\phi _{n}({\vec{r}}%
_{2})|^{2}/[(k+s_{n})^{2}+s_{n}^{\prime 2}]$, and ignoring the off-diagonal
terms, we get the transmission coefficient $T(k)$ expressed as a sum of
Lorentzian peaks 
\begin{equation}
T(k)\simeq \sum_{n}\frac{c_{n}}{(k-s_{n})^{2}+s_{n}^{\prime 2}}.
\label{transmission}
\end{equation}
The coupling $c_{n}$ in Eq. (\ref{transmission}) depend on the location of
the probes. The $k$ dependence of $c_{n}$ is locally weak near a resonance
since $1/[(k+s_{n})^{2}+s_{n}^{\prime 2}]$ is flat around $k\sim s_{n}\gg
s_{n}^{\prime }$ for sharp resonances. One can treat $c_{n}$ practically as
constants in the neighborhood of a resonance. This shows that the measured
transmission coefficient $T(k)$ shown in Fig. 1 directly yields the
wave-vector $s_{n}$ and the half-width $s_{n}^{\prime }$ of the sharper
scattering resonances. We next compare the experimental resonances with
semiclassical calculations.

\subsection{Calculation of Quantum Resonances}

Although any chaotic system can be quantized numerically by directly
diagonalizing the Hamiltonian truncated in certain expansion space, for a
large number of systems, one uses the techniques based upon semiclassical
periodic orbit theory \cite{Gutzwiller90}, such as the cycle expansion \cite
{Cvitanovic89}, Fredholm determinant \cite{Cvitanovic93b} or harmonic
inversion \cite{Main}. Except for the lowest eigenvalues, the semiclassical
quantization gives very accurate results. Using Gutzwiller's trace formula 
\cite{Gutzwiller90}, the semiclassical Ruelle zeta-function can be derived
as an {\it Euler} product over all the prime PO which are the PO without
repetition \cite{Cvitanovic93,Gaspard94}

\begin{equation}
\zeta _{j,sc}=\prod_{p}(1-t_{p,sc})^{-1},  \label{semi-Ruelle}
\end{equation}
with $t_{p,sc}$ the semiclassical weight

\begin{equation}
t_{p,sc}=(-1)^{l_{p}}\exp (ikL_{p})/|\Lambda _{p}|^{1/2}\Lambda _{p}^{j}.
\label{weight-semi}
\end{equation}
Here $l_{p}$ is the number of collisions of the PO with the disks, $L_{p}$
is the length of the PO and $\Lambda _{p}$ the eigenvalue of the instability
matrix ${\bf J}_{p}$. $j$ comes in from the decomposition of the trace of
the Green's function into Ruelle zeta-functions with $j=0,1,\cdots $. The
quantum resonances are the poles of the Ruelle zeta-functions which is
directly related to the trace of scattering matrix or the trace of the
Green's function. If a symbolic dynamics exists for the system, only a few
prime POs will be needed in the zeta-function to give quite accurate
eigenvalues because the curvature in the cycle expansion will decay
exponentially with increasing PO length \cite{Cvitanovic89}.

As an example, consider the non-chaotic $2$-disk system with disk separation 
$R$ and disk radius $a$. There is only one unstable prime PO between the
disks. The symmetry group of PO is $C_{2}$, with two one-dimensional
irreducible representations, symmetric $A_{1}$ and anti-symmetric $A_{2}$.
The semiclassical resonances in wave vector space are $k_{n}=[n\pi
+i(1/2)\ln \Lambda ]/(R-2a)$ with $\Lambda =\sigma -1+\sqrt{\sigma (\sigma
-2)}$ the eigenvalue of the instability matrix in the fundamental domain and 
$\sigma =R/a$. Here $n$ is odd for $A_{1}$ representation, $n$ even and $%
n\neq 0$ for $A_{2}$ representation \cite{Wirzba92}.

For the chaotic $n$-disk system with $n\geq 3$, there is no analytical
expression of the semiclassical quantum resonances. To do the numerical
calculations of the quantum resonances, use can be made of the symmetry of
the system and the cycle expansion in the fundamental 
domain \cite{Cvitanovic93,Gaspard94}. We have
calculated numerically the quantum resonances of the 3-disk and 4-disk
systems with different disk separation ratio $\sigma $ by searching the
poles of the Ruelle zeta function (\ref{semi-Ruelle}). An example of the
results of the calculation for the $s_{n},s_{n}^{\prime }$ for a 3-disk
system is shown in Fig. 2. The experiments only access a small range of the
quantum resonances with $0<%
\mathop{\rm Re}%
(f)<20GHz$ and $-0.3GHz<%
\mathop{\rm Im}%
(f)<0$. The experimental quantum resonances are well matched by the sharp
resonances from the semiclassical calculations. The very broad resonances
are not observed because the coupling to them is weak. For details
concerning the comparison of experimental and calculated quantum resonances,
see Ref. \cite{Lu00}.

\subsection{Experimental Spectral Autocorrelation{\em \ }$C(\kappa )$}

The spectral autocorrelation function is determined from the experimental
spectra as $C_{T}(\kappa )=\left\langle T(k-(\kappa /2))T(k+(\kappa
/2))\right\rangle _{k}$ with average carried out over a band of wave vector
centered at certain value $k_{0}$ and of width $\Delta k$ \cite{Lu00}. The
spectral autocorrelation can be fitted well by a superposition of
Lorentzians :

\begin{equation}
C_{T}(\kappa )=\sum_{\pm ,n=1}^{\infty }\frac{b_{n}\gamma _{n}^{\prime }}{%
\gamma _{n}^{\prime 2}+(\kappa \pm \gamma _{n}^{\prime \prime })^{2}},
\label{auto-cor}
\end{equation}
with $\gamma _{n}^{\prime }\pm i\gamma _{n}^{\prime \prime }$ the classical
RP resonances in wave vector space. A semiclassical derivation will be given
in Sec. 3.

The experimental RP resonances are obtained by fitting the experimental
autocorrelation as shown in Fig. 3 with Eq. (\ref{auto-cor}). Similar
comparisons between experiment and theory for the $n$-disk system with $%
n=2,3,4$ are presented in Ref. \cite{Pance00}. The coupling $b_{n}$ in the
above equation which should depend on the location of the probes determine
the decay probability of the classical RP resonances. Since we do not have
knowledge of them experimentally, they are chosen to optimize the fitting.

The experimental RP resonances ($\gamma _{n}^{\prime },\gamma _{n}^{\prime
\prime }$) obtained from the Lorentzian decomposition of the experimental $%
C(\kappa )$ as shown in Fig. \ref{fig-exp-cor} are displayed in Table 1.
They can be compared with theoretically calculated RP resonances, which are
the poles of the classical Ruelle zeta-function, and whose calculation will
be explained in the next subsection. Although the position of the peaks of
the oscillations in the experimental autocorrelation in Fig. \ref
{fig-exp-cor} is quite accurately given by the imaginary part $\gamma
_{n}^{\prime \prime }$ of the RP resonances, the experimental half-width of
the oscillation is almost always smaller than the calculated real part $%
\gamma _{n}^{\prime }$ of the RP resonances. This is possibly because the
absorber used is not ideal, and leads to quantum resonances that are
slightly sharper than calculated. This results in systematically sharper
widths of the experimental RP resonances.

\subsection{Calculation of Classical resonances}

The theoretical RP resonances are essential ingredients of a Liouvillian
description of the classical dynamics, and can be calculated as the poles of
the classical Ruelle zeta-function. The classical dynamics of Hamiltonian
flow in symplectic phase space can be described by the action of a linear
evolution operator which is also the Perron-Frobenius operator ${\cal L}%
^{t}(y,x)=\delta (y-f^{t}(x)).$ Here $f^{t}(x)$ is the trajectory of the
initial point $x$ in phase space. The trace of the Perron-Frobenius operator
is ${\rm tr}{\cal L}^{t}=\int \delta
(x-f^{t}(x))dx=\sum_{p}T_{p}\sum_{r=1}^{\infty }\delta (t-rT_{p})\left| \det
({\bf 1}-{\bf J}_{p}^{r})\right| ^{-1}$. The first summation is over all the
prime classical PO $p$ with period $T_{p}$, the second one is over the
repetition $r$ of the prime PO. ${\bf J}_{p}$ again is the instability
matrix. The evolution operator has a Lie group structure ${\cal L}%
^{t}=e^{-At}$ with $A$ the generator of the Hamiltonian flow. This generator 
$A$ has the classical resonances as its eigenvalues, $\gamma _{n}=\gamma
_{n}^{\prime }\pm i\gamma _{n}^{\prime \prime }$, which we call the RP
resonances \cite{Ruelle86,Pollicott}. We have ${\rm tr}{\cal L}%
^{t}=\sum_{n=0}^{\infty }g_{n}e^{-\gamma _{n}t},$ with $g_{n}$ the
multiplicity of the resonances. These resonances determine the time
evolution of any classical quantity. The trace formula for classical flows
is obtained from the Laplace transform of the above expression \cite
{Cvitanovic91} 
\begin{eqnarray}
{\rm tr}{\cal L}(s) \hskip -.3cm&=&\hskip -.3cm\int_{0}^{\infty }
dte^{st}{\rm tr}{\cal L}^{t}={\rm tr%
}(A-s{\bf 1)}^{-1}  \nonumber \\
\hskip -.3cm&=&\hskip -.3cm\sum_{p}T_{p}\sum_{r=1}^{\infty }
\frac{e^{rsT_{p}}}{\left| \det ({\bf 1}-%
{\bf J}_{p}^{r})\right| }.  \label{tra-class}
\end{eqnarray}

In order to calculate the RP resonances, the Ruelle zeta-function is
introduced as

\begin{equation}
\zeta _{\beta ,cl}=\prod_{p}(1-t_{p,cl})^{-1}.  \label{classical-Ruelle}
\end{equation}
Here the product is over all the prime PO with $t_{p,cl}$ the classical
weights of the periodic orbits 
\begin{equation}
t_{p,cl}=\exp (sT_{p})/|\Lambda _{p}|\Lambda _{p}^{\beta -1}.
\label{weight-classical}
\end{equation}
Here the integer $\beta >0$ comes in from the expansion of the determinant $%
\left| \det ({\bf 1}-{\bf J}_{p}^{r})\right| ^{-1}$. One can see that the
above classical Ruelle zeta-function is very similar to the semiclassical one (Eq.%
(\ref{semi-Ruelle})). The classical Ruelle zeta-function is exact and can be
directly derived from the above trace formula (\ref{tra-class}) of the
Perron-Frobenius operator. The topological pressure $P(\beta )$ can be
defined as the simple pole of the classical Ruelle zeta-function \cite
{Ruelle78} with $\beta $ extrapolated to the entire real space. All the
characteristic quantities of the classical dynamics, such as the
Kolmogorov-Sinai entropy, escape rate, fractal dimensions, can be derived
from $P(\beta )$ \cite{Gaspard92}. For the hard-disk system, the classical
velocity $\upsilon $ of particles is constant. So $T_{p}=L_{p}/\upsilon $.
For simplicity, one can set $\upsilon =1$. The RP resonances are then
calculated in wave-vector space.

For example, the classical RP resonances of the 2-disk system are $\gamma
_{n}=[\ln \Lambda +in\pi ]/(R-2a)$ with $\Lambda $ given before. Here $n$ is
even for $A_{1}$ representation and $n$ odd for $A_{2}$ representation.
Similar to the situation of the quantum resonances $s_{n}+is_{n}^{\prime }$,
there is no analytical expression of the RP resonances $\gamma _{n}^{\prime
}\pm i\gamma _{n}^{\prime \prime }$ for the chaotic $n$-disk system. They
can be calculated numerically as the poles of the classical Ruelle
zeta-function Eq. (\ref{classical-Ruelle}) \cite{Gaspard92}. 14 prime PO up
to period 3 were used in our calculation of the RP resonances of the 4-disk
system, the results of which are shown in Table 1.

\subsection{Numerical simulation}

To gain further insight into the established experimental correspondence, we
have utilized the results of the calculations of the quantum and classical
resonances to carry out a numerical simulation of the physical experiment.
We calculate numerically a close approximation to the experimental quantum
spectrum using the quantum resonances. We them determine the spectral
autocorrelation of the simulated spectrum, and compare its decomposition
with the calculated classical resonances.

Writing Eq. (\ref{transmission}) explicitly, we have following form of the
transmission $T(k)\simeq (2m/\hbar ^{2})|A(k)|^{2}\sum_{n}\frac{|\psi _{n}({%
\vec{r}}_{1})\phi _{n}({\vec{r}}_{2})|^{2}}{\left[
(k+s_{n})^{2}+s_{n}^{\prime 2}\right] \left[ (k-s_{n})^{2}+s_{n}^{\prime
2}\right] }$. Here we assume the presence of the probes poses a small
perturbation and will not change the quantum spectrum. For a closed chaotic
system with time-reversal symmetry, the wave density is expected to follow
the Porter-Thomas (PT) distribution in RMT \cite{Kudrolli95}. For a very
open system and when the probes are far away from the scatterers, the wave
density will not follow the PT distribution. But in the vicinity of the
scatterers, one may assume the density to follow the PT distribution.
Setting $\rho _{1n}=|\psi _{n}({\vec{r}}_{1})|^{2}$ and $\rho _{2n}=|\phi
_{n}({\vec{r}}_{2})|^{2}$, one has, $P(\rho _{n})=\left( 2\pi \rho _{n}%
\overline{\rho }\right) ^{-1/2}\exp (-\rho _{n}/2\overline{\rho })$ with $%
\overline{\rho }$ being the ensemble averaged density. The densities at the
location of different antennas were found not to be correlated in closed
cavities \cite{Alt95}. We assume the same is true for an open chaotic system in
the vicinity of the scatterer disks. For simplicity, we consider the
ensemble average of the transmission coefficient $\overline{T(k)}\simeq
(2m/\hbar ^{2})|A(k)|^{2}\overline{\rho }_{1}\overline{\rho }%
_{2}\sum_{n}1/\left[ (k+s_{n})^{2}+s_{n}^{\prime 2}\right] \left[
(k-s_{n})^{2}+s_{n}^{\prime 2}\right] $. From this expression, one can see
that the contribution from broad resonances is suppressed. The main
contribution is from sharp resonances. This has already been observed in our
experiments\cite{Lu99,Lu00}. For these sharp resonances, if they are far
away from the origin, $(k+s_{n})^{2}+s_{n}^{\prime 2}\simeq 4k^{2}$. Thus, for
large $k\gg s_{n}^{\prime }$, the above transmission can be approximated as $%
\overline{T(k)}\simeq (2m/\hbar ^{2})|A(k)|^{2}(\overline{\rho }_{1}%
\overline{\rho }_{2}/4k^{2})\sum_{n}\left[ (k-s_{n})^{2}+s_{n}^{\prime
2}\right] ^{-1}$. The function $A(k)$ is found to be proportional to $k$ 
\cite{Stein95,Stockmann99}. For convenience, we set $A(k)=2k$ and also $%
(2m/\hbar ^{2})\overline{\rho }_{1n}\overline{\rho }_{2n}=1$. We thus get 
\begin{equation}
\overline{T(k)}\simeq \sum_{n}\frac{1}{\left[ (k-s_{n})^{2}+s_{n}^{\prime
2}\right] }.  \label{numeric}
\end{equation}
The $k$ dependence of $A(k)$ can be understand as follows. In the
experimental setup, one of the antenna can be regarded as a diploe radiating
electromagnetic waves. The radiated electrical field of a dipole is
proportional to $k$ if an alternating current with constant amplitude 
$I_{\rm in}$ was maintained on it \cite{Jackson}. The voltage picked up on
another antenna is just the electrical field at the antenna location times
the length $d$ of the antenna inside the cavity. The transmission amplitude $%
S_{21}$ measured by the analyzier is the ratio of the output voltage on one
antenna to the input current of another antenna, 
$S_{21}=dE_{z,{\rm out}}/I_{\rm in}$, 
thus $A(k)=dE_{z,{\rm in}}/I_{\rm in}\propto k$.

The expression (\ref{numeric}) is thus used to get the averaged transmission
and then to calculate the autocorrelation $C_{T}(\kappa )$ numerically. We
have done that for the $n$-disk system with $n=2,3,4$ using $a=5$cm. Here we
just present the results for the $3$-disk system as shown in Fig. 4 and 5.
Note that $f=(c/2\pi )k$, $1GHz$ is equivalent to $0.2096$cm$^{-1}$. The
symmetry group for the chaotic 3-disk system is $C_{3v}$ \cite{Cvitanovic93}%
, which has two one-dimensional irreducible representations $A_{1}$, $A_{2}$%
, and one two-dimensional irreducible representation $E$. In the fundamental
domain, only the resonances of the $A_{2}$ representation will contribute to
the transmission \cite{Lu99}. The semiclassical resonances of $A_{2}$
representation in the range $0<%
\mathop{\rm Re}%
ka<100$ and $-0.5<%
\mathop{\rm Im}%
ka<0$ for the disk separation $R/a=4\sqrt{3}$ are obtained from $\zeta
_{sc}(ik)$ with $j=0$. About 200 resonances are obtained as shown in Fig. 2.

The corresponding classical resonances shown in Fig. 5 are from the $A_{1}$
representation of the classical Ruelle zeta-function (\ref{classical-Ruelle}%
) with $\beta =1$. 8 prime PO up to period 4 were used in our calculations
of the quantum and RP resonances. The numerical transmission and the
autocorrelation are shown in Fig. 4 and 5, respectively. One can see that
the oscillations in the autocorrelation are fully determined by the
classical RP resonances as predicted by Eq.(\ref{auto-cor}).

\section{Quantum and classical correlations}

In the above section, we have demonstrated experimentally and numerically
that the quantum auto-correlation and hence the statistics of quantum
resonances are determined by the RP resonances of the underlying classical
dynamics. Here we explore the theoretical justification for this
correspondence.

For chaotic open scattering systems, the distribution and correlation of the
quantum resonances have been discussed in the framework of RMT \cite
{Fyodorov98}. However, the RMT results are the ensemble averages of
different systems with certain symmetry in common. Consequently, RMT does
not make contact with system specific features of the underlying classical
dynamics. Thus it is unable to make any concrete connection between the
quantum and classical dynamics, for which it is necessary to go beyond RMT.

The appearance of nonuniversality requires the use of structures such as
periodic orbits within semiclassical calculations. To proceed, we first
consider the correlation of the density of states. In the semiclassical
periodic orbit theory, the fluctuation part of the density of states is \cite
{Gutzwiller90}

\begin{eqnarray}
{\cal D}(k) \hskip -.3cm&=&\hskip -.3cm-\frac{1}{\pi }%
\mathop{\rm Im}%
{\rm tr}G({\vec{r}}_{1},{\vec{r}}_{2},k)  \nonumber \\
\hskip -.3cm&=&\hskip -.3cm-\frac{1}{\pi }%
\mathop{\rm Im}%
\sum_{p}T_{p}\sum_{r=1}^{\infty }\frac{e^{irkL_{p}}}{\left| \det ({\bf 1}-%
{\bf J}_{p}^{r})\right| ^{1/2}}.  \label{density}
\end{eqnarray}
The trace of the Green's function is the integration over the configuration
space with ${\vec{r}}_{1}={\vec{r}}_{2}$. The autocorrelation is

\[
C_{{\cal D}}(\kappa )=\frac{1}{2K}\int_{-K}^{K}dk{\cal D}(k+\frac{\kappa }{2}%
){\cal D}(k-\frac{\kappa }{2}),\ \;\;K\gg \kappa . 
\]
Substituting the semiclassical expression (\ref{density}) of ${\cal D}(k)$
into the above equation, one gets \cite{Cvitanovic00}

\begin{eqnarray}
C_{{\cal D}}(\kappa ) \hskip -.3cm&=&\hskip -.3cm
\mathop{\rm Re}%
\sum_{p}T_{p}^{2}\sum_{r=1}^{\infty }\frac{e^{rsT_{p}}}{\left| \det ({\bf 1}-%
{\bf J}_{p}^{r})\right| }\simeq 
\mathop{\rm Re}%
{\rm tr}\frac{1}{A-s{\bf 1}}  \nonumber \\
\hskip -.3cm&=&\hskip -.3cm\sum_{n}\frac{1}{\upsilon }\left[ \frac{\gamma _{n}^{\prime }}{\gamma
_{n}^{\prime 2}+(\kappa +\gamma _{n}^{\prime \prime })^{2}}+\frac{\gamma
_{n}^{\prime }}{\gamma _{n}^{\prime 2}+(\kappa -\gamma _{n}^{\prime \prime
})^{2}}\right]  \label{semi-cor}
\end{eqnarray}
with $s=i\upsilon \kappa $. Since the spectrum of the generator $A$ is $%
\upsilon (\gamma _{n}^{\prime }\pm i\gamma _{n}^{\prime \prime })$, one can
see that in the semiclassical theory, the autocorrelation of the density of
states is determined solely by the classical RP resonances. The transmission
spectrum $T(k)\,$is a projection of the density of states, one expects that
the autocorrelation of the transmission is also determined by the RP
resonances. The approximation in Eq. (\ref{semi-cor}) is valid only for
hyperbolic systems with finite symbolic dynamics. For these systems as
explained in Sec. 2.1, the contribution to the autocorrelation $C_{{\cal
D}%
}(\kappa )$ is mainly from a few fundamental prime POs. For those PO, one
has $T_{p}^{2}\simeq \overline{T}T_{p}$ with $\overline{T}$ the average
period. For weakly open or closed system, the approximation will break down.

The experimental quantity whose correlations we examined in Sec. 2 is the
Green's function, while theoretical arguments usually examine the density of
states or the time delay \cite{Eckhardt93} which are however not directly
measurable experimentally. To develop the correlation theory for the
transmission we measured, consider the stationary Green's function $G({\vec{r%
}},{\vec{r}}_{0},k)$ which is the solution of the following equation $%
(\nabla ^{2}+k^{2})G({\vec{r}},{\vec{r}}_{0},k)=\delta ({\vec{r}}-{\vec{r}}%
_{0}),$ with certain boundary condition. The quantum mechanical propagator
is the Fourier transform of the Green's function. One can construct a time
evolution of a wave packet ${\hat{K}}({\vec{r}},{\vec{r}}_{0},t)${\ }with ${%
\hat{K}}({\vec{r}},{\vec{r}}_{0},t)=(2\pi \hbar i)^{-1}\int_{\Delta
\varepsilon }G({\vec{r}},{\vec{r}}_{0},\sqrt{2m\varepsilon /\hbar ^{2}}%
)e^{-i\varepsilon t/\hbar }d\varepsilon ,$ with $\varepsilon =\hbar
^{2}k^{2}/2m$. Here the integration is performed around $\varepsilon
_{0}=\hbar ^{2}k_{0}^{2}/2m$, with the range $\Delta \varepsilon =\hbar
\upsilon \Delta k$ and $\upsilon =\hbar k_{0}/m$ the group velocity of the
wave packet. The integration in the $\varepsilon $ space can be changed into
that in the $k$ space. We get ${\hat{K}}({\vec{r}},{\vec{r}}%
_{0},t)=e^{-i\varepsilon _{0}t/\hbar }(\upsilon /2\pi i)\int_{\Delta k}G({%
\vec{r}},{\vec{r}}_{0},k_{0}+k)e^{-i\upsilon kt}dk.$ The propagator ${\hat{K}%
}({\vec{r}},{\vec{r}}_{0},t)$ is just the wave function at point ${\vec{r}}$
due to a $\delta $-function excitation of the system at point ${\vec{r}}_{0}$
and time $t_{0}=0$. The particle density $\rho (t)$ is thus 
\begin{eqnarray*}
\rho (t) \hskip -.3cm&=&\hskip -.3cm\left| {\hat{K}}({\vec{r}},{\vec{r}}_{0},t)\right| ^{2}=\frac{%
\upsilon ^{2}}{4\pi ^{2}}\int_{\Delta k}G({\vec{r}},{\vec{r}}%
_{0},k_{0}+k)e^{-i\upsilon kt}dk \\
\hskip -.3cm&&\hskip -.3cm\times \int_{\Delta k}G^{*}({\vec{r}},{\vec{r}}_{0},k_{0}+k^{\prime
})e^{i\upsilon k^{\prime }t}dk^{\prime },
\end{eqnarray*}
with the average given by $\left\langle \rho (t)\right\rangle
_{t}=\lim_{T\to \infty }{T}^{-1}\int_{0}^{T}\rho (t)dt=({\upsilon ^{2}/}2\pi
L)\int dk\left| G(k)\right| ^{2}.$ A probability measure is assumed for the
above average to converge \cite{Ruelle86}. Here $L$ is the size of the whole
system. For an open system, $L\to \infty $. Here we used the definition of
the $\delta $ function $\delta (t)=(2\pi )^{-1}\int d\omega e^{-i\omega t}.$

The autocorrelation of the particle density is 
\[
C_{\rho }(\tau )=\left\langle \rho (t)\rho (t+\tau )\right\rangle
_{t}-\left\langle \rho \right\rangle _{t}^{2}. 
\]
Using the diagonal approximation, we get 
\begin{equation}
C_{\rho }(\tau )={\frac{\upsilon ^{4}\Delta k}{4\pi ^{2}L^{2}}}\int d\kappa
C_{T}(\kappa )e^{-i\upsilon \kappa \tau }.  \label{fourier}
\end{equation}
Note that we have $T(k)\propto \left| G({\vec{r}},{\vec{r}}_{0},k)\right|
^{2}.$

If one assumes that the correlation of the particle density is classical,
one has \cite{Chris90} 
\begin{equation}
C_{\rho }(\tau )=\sum_{n=0}^{\infty }2b_{n}^{\prime }e^{-\gamma _{n}^{\prime
}\upsilon \tau }\cos \gamma _{n}^{\prime \prime }\upsilon \tau ,
\label{decay-mode}
\end{equation}
where the coefficients $b_{n}^{\prime }$ are the coupling, and $\gamma
_{n}^{\prime }\pm i\gamma _{n}^{\prime \prime }$ the RP resonances evaluated
in wave-vector space. The Fourier transformation of (\ref{fourier}) and (\ref
{decay-mode}) gives 
\begin{equation}
C_{T}(\kappa )=\sum_{\pm ,n=0}^{\infty }\frac{b_{n}\gamma _{n}^{\prime }}{%
\gamma _{n}^{\prime 2}+(\kappa \pm \gamma _{n}^{\prime \prime })^{2}}.
\end{equation}
with $b_{n}=2\pi L^{2}b_{n}^{\prime }/\upsilon ^{4}\Delta k$. The
autocorrelation is usually normalized as $C_{T}(0)=1$ so that the system
size $L$ will be cancelled. Thus the above expression is valid for both
closed and open systems.

The above arguments closely parallel Agam's \cite{Agam00} recent application
of the diagrammatic techniques for disordered systems to open chaotic
systems. We remark that the above arguments differ somewhat from those of
Agam, whose results are valid when the two probes are far away from each
other which implies $\left\langle T^{2}\right\rangle =2\left\langle
T\right\rangle ^{2}$. This is the case in the diffusive limit with a Poisson
distribution of the transmission coefficient $T$.

The above correspondence between the autocorrelation derived from
semiclassical theory and the classical resonances is obviously not exact,
but requires corrections due to quantum interference. These corrections are
small for open systems, but are expected to dominate for closed systems.

\section{Discussion and concluding remarks}

The results presented here establish an alternative path from quantum to
classical achieved by considering the correspondence of quantum correlations
and classical decay modes. As Fig. \ref{fig-exp-cor} shows, the small $%
\kappa $ behavior of $C(\kappa )\,$is universal, while the large $\kappa $
behavior shows non-universal oscillations that are completely described by
the classical RP resonances. This demonstrates that in an open system,
quantum correlation functions are intimately related to decay laws of
classical survival probabilities or evolution of particle ensembles.

Most theoretical calculations in quantum chaos using ensemble averaging are
restricted only to the leading order consistent with RMT, which probes only
the average properties of the system, such as the average escape rate $%
\gamma $. The leading RMT contribution to the correlation $C(\kappa )$ is
Lorentzian \cite{Blumel88,Jalabert90,Lewenkopf91,Lai92,Lu99} $C(\kappa
)=C(0)/(1+(\kappa /\gamma _{cl})^{2})$, when the classical decay is
exponential, which occurs for hyperbolic systems such as the $n$-disk
billiards. Our earlier experiments established this result in the open
microwave billiards \cite{Lu99}.

As we have shown, our experiments go well beyond the average information
represented by RMT - the spectral data contain further details of the fine
structure embodied in the decay modes or classical resonances of the system.
In terms of the fractal repeller, the wave mechanical experiments thus yield
not only the leading average property $\gamma _{cl}$ related to the fractal
dimension, but also further higher order properties of the repeller in terms
of the classical resonances. While we have demonstrated this for the $n$%
-disk billiard, the above result should be valid generally for hyperbolic
open chaotic systems.

The relevant classical limit here is the evolution of classical particle
ensembles. The notion of ensemble averaging that is common in quantum
mechanics is also useful in classical treatments of chaotic systems. This is
because while the prediction of individual particles is sensitive to initial
conditions, the dynamics of appropriate averages of a perturbed system
relaxing towards equilibrium is well defined. The RP resonances determine
the decay modes of the dynamical systems evolving from a nonequilibrium
state to equilibrium or steady state. For an open system, Gaspard and
Nicolis \cite{Gaspard90} have derived an escape rate formula to relate the
coefficient of diffusion and the escape rate of large open systems. While
for an open system, the escape rate $\gamma _{0}$ is nonvanishing, for
closed system, $\gamma _{0}=0$, and the corresponding eigenstate is just the
equilibrium state. The RP resonances are essential ingredients of a
description of irreversibility and thermodynamics based upon microscopic
chaos \cite{Hasegawa92,Gaspard92,Dorfman99}. However these resonances have
remained mathematical objects. The present experiments have provided
physical reality to them. It is remarkable that we are able to observe
classical resonances in a quantum experiment. Because microwaves do not
interact (collide) with each other, ensemble averaging is easier to achieve
in microwaves than in fluids. Thus theories that employ ensemble averaging
are ideally applicable to the microwave experiments. This is true for both
open systems as shown here, and for disordered billiards \cite{Pradhan00}.

The present work shows that the information concerning the classical
resonances is coded into the quantum spectrum. This is somewhat anticipated
in periodic orbit theory, where one can observe the strong similarity
between the classical (Eq.(\ref{classical-Ruelle}) and (\ref
{weight-classical})) and semiclassical (Eq.(\ref{semi-Ruelle}) and (\ref
{weight-semi})) Ruelle zeta-functions. A one-to-one correspondence exists
between $k_{n}$ and $\gamma _{n}$ for the $2$-disk system as discussed in
Sec.I. Another example where an exact relation between the classical and
quantum resonances exists is the system of a free particle sliding on a
compact surface of constant negative curvature, the RP resonances were found
by Biswas and Sinha \cite{Biswas93} to be $\gamma _{0}=0$, $\gamma _{n}=%
\frac{1}{2}\pm ip_{n}$ for $n\geq 1$ with real $p_{n}$ the solution of a
certain equation while the quantal excitation eigenvalues are $%
E_{n}=p_{n}^{2}+\frac{1}{4}$. One would also expect a relation between the
distribution of the classical and semiclassical resonances. The RP
resonances have recently been observed in the distribution of the zeroes of
the Riemann zeta function \cite{Bohigas00}.

Our work demonstrates that suitable quantum correlations diffuse just like
classical observables in an open system. An interesting question is what
will happen when the number $n$ of disks are very large, a situation that is
frequently referred to as the Lorentz gas. Due to the exponential separation
of the nearby orbits, a particle is forced to move through the bouncing
paths, or undergo diffusive motion, for long time. One may argue that the
particle motion occurs in a system of size $L${\bf , }in which it has a
diffusive path and hence the average escape time $t_{c}$ $\sim $ $L^{2}/D$,
with $D$ the diffusion coefficient. Thus the escape rate $\gamma \sim $ $%
D/L^{2}$. A more general expression for diffusion coefficient $D$ is derived
through the escape rate formula in Ref. \cite{Gaspard90}. For the Lorentz
gas with large $L$, by taking the general form of the escape rate $\gamma
=\lambda -h_{ks},$ where $\lambda $ is the Lyapunov exponent and $h_{KS}$ is
the Kolmogorov-Sinai entropy of the fractal repeller ${\cal F}_{L}$, one has 
$D=\left[ L/2.40482\right] ^{2}[\lambda ({\cal F}_{L})-h_{KS}({\cal F}_{L})]$.

There is a fundamental connection between the present observation of
classical resonances and the diffusion operator approach of the
supersymmetry theories \cite{Efetov} of disordered systems. The statistics
of the spectra and eigenfunction of disordered systems is directly related
to the classical spectral determinant $D(\varepsilon )=\varepsilon
^{-2}\prod_{\mu }(1+\varepsilon ^{2}\Delta ^{2}/\gamma _{\mu }^{2})^{-1}$,
where $\varepsilon $ is the energy ( in unit of $\Delta $, the average level
spacing), and $\gamma _{\mu }$ are the non-zero eigenvalues of the diffusion
operator \cite{Andreev95,Mirlin00}. These are just the RP resonances that we
have observed in open systems. Using the supersymmetric techniques, Agam,
Altshuler, and Andreev \cite{Agam95} calculated the level correlation and
showed that it can be fully determined by the classical spectrum of the
Perron-Frobenius operator. Bogomolny and Keating \cite{Bogomolny96} arrived
at a similar conclusion using the periodic orbit approach. The applicability
of these theories for closed systems is currently a matter of debate \cite
{Prange97}, but the applicability of the supersymmetry theory to disordered
systems is well established, and is strongly supported by microwave
experiments on disordered billiards \cite{Kudrolli95,Pradhan00}.

As pointed out by Berry \cite{Berry91}, the two limits $T\to \infty $ and $%
\hbar \to 0$ do not commute with each other, thus the long-time quantum
evolution is fundamentally different from long time classical evolution. The
probability $P_{q}$ that the quantum chaotic system decays at time $t$ after
its formation follows a power law \cite{Harney92,Alt95} $P_{q}(t)\sim (1+2%
\overline{\Gamma }t)^{-2-\frac{M}{2}},$ for $t$ larger than the equilibrium
time of the system. Here $\overline{\Gamma }$ is the average decay width, $M$
the number of open channels. As the system is opened up with more channels,
the decay will be almost exponential since the above equation can be
approximated by $P_{q}(t)\sim \exp \left[ -(4+M)\overline{\Gamma }t\right] ,$
for large $M$ and the weight of the algebraic tail becomes negligible. More
precisely, as pointed out in Ref. \cite{Casati97}, there exists a new
quantum relaxation time scale $t_{q}<t_{H}$, with $t_{H}=\hbar /\Delta $ the
Heisenberg time. Beyond that time, the quantum correlation decays
algebraically. For shorter times, the quantum correlation decays
exponentially in the same way as the classical correlation does. For an open
system, since the spectrum is continuous, the Heisenberg time is actually
infinite, and so is the quantum relaxation time $t_{q}$. This is one reason
that the classical resonances are clearly visible in the present
experiments, and the Lorentzian decomposition of the spectral
autocorrelation works so well.

The transmission spectrum measured in the microwave experiment directly
corresponds to the conductivity measured in electronic quantum dots. In
quantum dot experiments though, the magnetoconductivity $\sigma (B)$ is
typically measured, so that the resulting autocorrelation $C(\Delta B)$ is
related to the area spectrum, rather than the wave vector dependence $%
C(\kappa )$ measured here which is related to the length spectrum \cite
{Jalabert90}. Nevertheless similar arguments should be applicable there
also, as well as to a variety of other physical systems which are
essentially open and chaotic, such as molecular photo-dissociation \cite
{Agam99} and chemical reactions \cite{Gaspard89}.

While the coupling to the quantum resonances is well understood, the
statistics and also the physical meaning of the experimental coupling $b_{n}$
to the classical resonances in Eq. (\ref{auto-cor}) deserves further
clarification. It would also be very interesting to examine the role of
classical resonances in mixed phase space systems \cite{Seba00}, 
such as the open billiards frequently considered in quantum dots 
where the classical decay is non-exponential \cite{Ketzmerick96}. 
More detailed work should be devoted to these and other interesting
questions raised by the present results.

\section*{Acknowledgements}

We have benefited greatly from formal and informal discussions at the
``Quantum Chaos Y2K'' Nobel Symposium. We thank P. Cvitanovi\'{c} and O.
Agam for important insights into the semiclassical arguments for the
correspondence between quantum and classical correlations. We also thank O.
Bohigas, B. Altshuler, J. Keating, E. Bogomolny, D. Ullmo and P. Lebeouf for
useful discussions.

This work was supported by NSF-PHY-9722681.

\begin{footnotesize}


\end{footnotesize}

\begin{table}[htbp] \centering%
\caption{\footnotesize Experimental and theoretical RP resonances of the 4-disk
system in the fundamental domain with $R/a=4\sqrt{2}$ and $a=5$ cm.
The RP resonances are in the unit of cm$^{-1}$.
\label{key}} 
\vskip .4cm
\begin{tabular}{llll}
\hline \hline
$\gamma _{\exp }^{\prime \prime }$ & $\gamma _{\rm{th}}^{\prime \prime }$
& $\gamma _{\exp }^{\prime }$ & $\gamma _{\rm{th}}^{\prime }$ \\ 
0.000 & 0.000 & 0.049 & 0.0599 \\ 
0.187 & 0.242 & 0.053 & 0.0980 \\ 
0.378 & 0.386 & 0.051 & 0.0910 \\ 
0.515 & 0.468 & 0.165 & 0.1682 \\ 
0.632 & 0.629 & 0.079 & 0.0647 \\ 
0.842 & 0.868 & 0.055 & 0.0841 \\ 
1.016 & 1.027 & 0.045 & 0.0785 \\ 
1.060 & 1.125 & 0.096 & 0.1067 \\ 
1.254 & 1.255 & 0.060 & 0.0718
\\\hline \hline
\end{tabular}
\end{table}%

\begin{figure}[htbp]
\epsfig{file=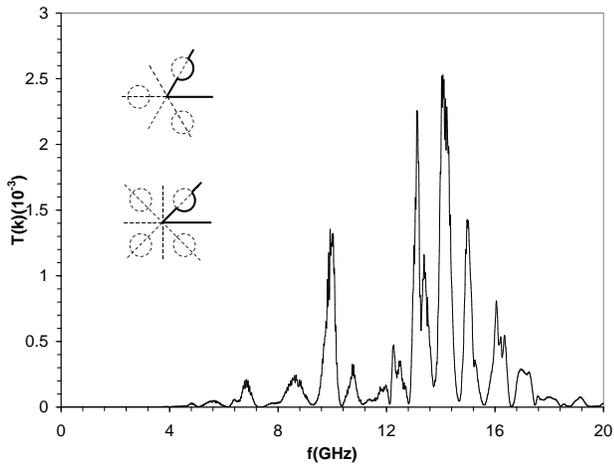,width=2.4in,angle=-90}
\caption{\footnotesize Experimental
transmission $T(k )$ of the $4$-disk system in the fundamental
domain  with $R/a=4\sqrt{2}$ and  $a=5$ cm. Here $f=(c/2\pi)k$. Inset:
Geometry of the 3- and 4-disk billard. 
The solid lines represent the fundamental domain in which the experiments were carried out. }
\label{tra-exp}
\end{figure}%

\begin{figure}[htbp]
\epsfig{file=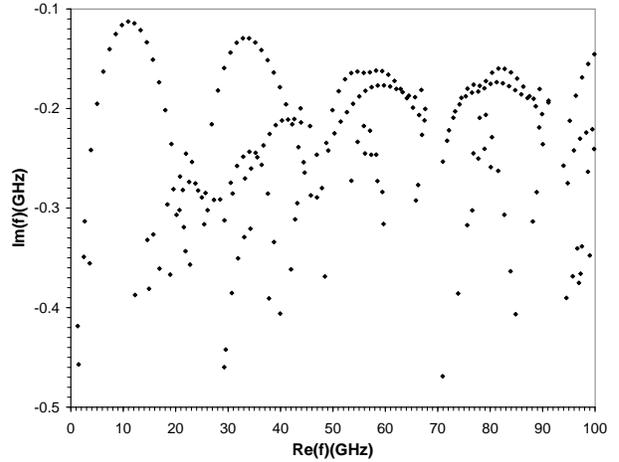,width=3.2in,angle=0}
\caption{\footnotesize Quantum
resonances in the $ A_2$ representation of the 3-disk system with 
$R/a=4\sqrt{3}$ with $a=5$ cm  in the fundamental domain. The quantum resonances we
observed experimentally are in the range $0<f<20GHz$. See \cite{Lu00} for details. }
\label{quan-res}
\end{figure}%

\begin{figure}[htbp]
\epsfig{file=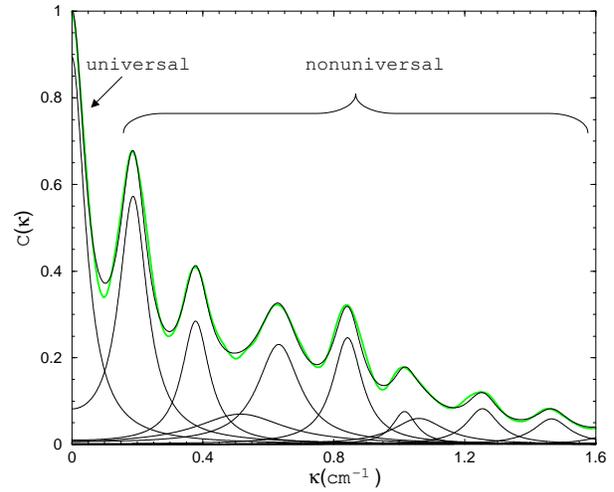,width=2.6in,angle=-90}
\caption{\footnotesize Autocorrelation
function $C(\kappa )$ vs $\kappa$ 
(cm$^{-1}$) of the $4$-disk system in the fundamental
domain  with $R/a=4\sqrt{2}$ and  $a=5$ cm.
 The gray line is the experimental $C(\kappa)$  calculated from the experimental trace shown
in Fig. 1. 
The data show the small $\kappa$ universal behavior followed by the non-universal oscillations
for large $\kappa$. The entire autocorrelation $C(\kappa )$ can be described in terms of the
RP resonances using Eq.(4) (thick solid line). The thin lines represent the Lorentzian decomposition
into individual RP resonances.  }
\label{fig-exp-cor}
\end{figure}%

\begin{figure}[htbp]
\epsfig{file=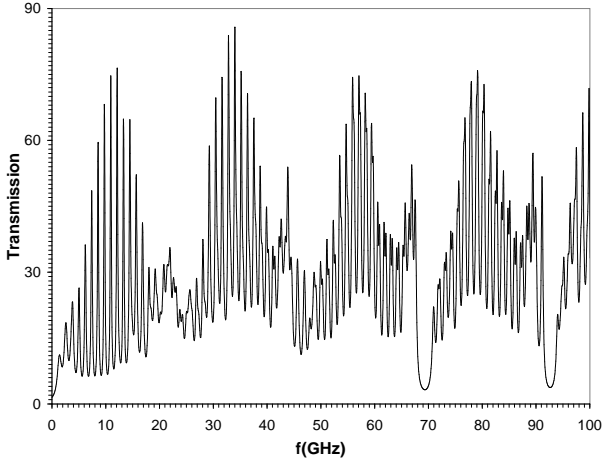,width=3.2in,angle=0}
\caption{\footnotesize Numerical
spectral transmission $\overline {T(k)}$ in the fundamental domain 
of the 3-disk system for $R/a=4\sqrt{3}$ and $a=5$ cm.}
\label{num-tra}
\end{figure}%

\begin{figure}[htbp]
\epsfig{file=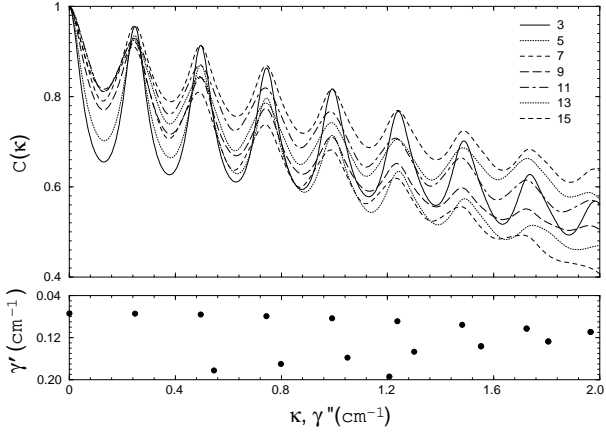,width=2.3in,angle=-90}
\caption{\footnotesize Numerical
spectral autocorrelation $C(\kappa)$ (top) and RP resonances
$\gamma'\pm i\gamma''$  (bottom)
 for the 3-disk system with 
$R/a=4\sqrt{3}$ and $a=5$cm in the fundamental domain. $C(\kappa)$ was calculated
from the numerical spectrum shown in Fig. 4  for interval
 $\Delta k=6$cm$^{-1}$ with the central values $k_0$ in cm$^{-1}$ indicated in the figure.}
\label{fig-3d-simu-cor}
\end{figure}%

\end{small}

\end{document}